\documentclass[english,aps,a4paper,twocolumn,showpacs,amsfonts,amsmath,amssymb,nofootinbib,notitlepage]{revtex4}
\usepackage[english]{babel}
\usepackage{graphicx}
\usepackage[dvips]{color}

\def\Journal#1#2#3#4#5#6{{#1}, {\it #4} \textbf{#5}, #6 (#2).}
\def\Book#1#2#3#4#5{{#1}, {\it #3} (#4, #5, #2).}

\newcommand{\PP}{\mathbb{P}}

\newcommand{\x}{\xi}
\newcommand{\y}{\upsilon}
\newcommand{\z}{\zeta}

\begin{document}

\title{Parrondo-like behavior in continuous-time random walks with memory}

\author{Miquel Montero}\email[E-mail: ]{miquel.montero@ub.edu}
\affiliation{Departament de F\'{\i}sica Fonamental, Universitat de
Barcelona (UB), Mart\'{\i} i Franqu\`es 1, E-08028 Barcelona, Spain}

\begin{abstract}
The Continuous-Time Random Walk (CTRW) formalism can be adapted to encompass stochastic processes with memory. In this article we will show how the random combination of two different unbiased CTRWs can give raise to a process with clear drift, if one of them is a CTRW with memory. If one identifies the other one as noise, the effect can be thought as a kind of stochastic resonance. The ultimate origin of this phenomenon is the same of the Parrondo's paradox in game theory.
\end{abstract}

\date{\today}
\maketitle

\section{Introduction}
\label{Sec_Intro}

The Continuous-Time Random Walk (CTRW) is the natural generalization of the discrete-time random walk: a stochastic process that shows changes of random magnitude at random (rather than fixed) instants of time. 
Since their introduction in 1965 by Montroll and Weiss in the physics literature~\cite{MW65,W94}, CTRWs have stood out for their versatility in the description of the random dynamics of a wide variety of systems. A quick review of the bibliography reveals applications in fields as diverse as: transport in heterogeneous media~\cite{S74,MS84,WPM98,MaBe02}, anomalous relaxation in polymer chains~\cite{HWS82}, electron tunneling \cite{GW02}, self-organized criticality \cite{BC97}, earthquake modeling~\cite{HS02,MAGLP03,C06}, random networks \cite{BS97}, transmission tomography~\cite{DWG03,DW05}, hydrology~\cite{BKMS01,DB03}, and tick-by-tick finance~\cite{SGM00,MMW03,KS03,RR04,MMP05,MPMLMM05,S06,MMPW06}. 

In the most ubiquitous version of the CTRW formalism~\cite{GPR01,K02,SGM04,GPSL09} the magnitudes of the steps (or jumps) and the time intervals between them (also called sojourns) constitute a two-dimensional set of independent and identically distributed (i.i.d.) random variables. While in many cases this is a convenient assumption, there are also examples in which correlations between consecutive step sizes and/or waiting times must be compulsorily considered. For instance, it has been reported the dependence that shows the seismic recurrence time on the magnitude of the last earthquake~\cite{C06}, or in the field of quantitative finance (to which the author has devoted a significant amount of his previous scientific activity) it is well known~\cite{C01} that the i.i.d. assumption is no longer valid when the market is observed  at the {\it atomic level\/}: At this scale one typically finds that price changes are negatively correlated~\cite{MPMLMM05,GK10}.

A plausible mechanism that (at least in part) explains the presence of this correlation is the so-called ``bid-ask bounce." Financial markets are normally double-auction markets where potential buyers and sellers simultaneously submit their bid and ask prices, the {\it limit orders\/}, to the market. In the most common situation, limit orders do not match previous limit orders (bid prices are lower than ask prices) and thus they are not executed, but recorded in the limit-order book. In fact, transactions are completed usually only after a {\it market order\/} (an order for buying or selling at the best available price) is introduced into the market. The random alternation in the arrival of buy and sell market orders makes that the last traded price bounces back and forth from the cheapest demand price to the highest offer price of the book. Thus, as long as bid and ask prices do not vary, consecutive price changes will show negative correlation.

All these empirical evidences encouraged the development of a new class of CTRWs based on the premise that the size of jumps and sojourns should depend on the previous values of these magnitudes~\cite{MM07}. The precise way in which the memory is introduced into the formalism is by demanding that the process {\it increments\/} satisfy the Markov property, what renders the problem tractable. In general, these processes can be easily connected with the broad family of the Markovian renewal processes~\cite{CM65} and, in particular, the CTRW with memory that we are going to analyze here is very alike to a
Markov chain with rewards~\cite{AAP05},~\footnote{Two main aspects distinguish CTRWs with memory from Markov renewal processes or Markov chains with rewards. In the first place, our jump sizes may take continuous values whereas the state space of a Markov chain is usually finite. In the second place, here the Markov property is satisfied not by the process but by the random jumps. However, in the case that memory does not affect the random occurrence of the time changes, the problem may recover the Markovian nature just by increasing the dimension of the process.}
a random game with heterogeneous payouts that may exhibit the {\it Parrondo's effect\/}.

The Parrondo's effect or paradox~\cite{HA99} is a counterintuitive feature that appears when two negatively biased (losing) games are combined to produce a positively biased (winning) game. This sort of games, first devised by J. M. R. Parrondo, has played a very relevant role in understanding the intriguing behavior shown by many physical systems, wherein the addition of disorder can lead to the emergence of some kind of order. This is the case of Brownian-ratchet related problems~\cite{HAT00,AATA04}, but Parrondo's games may have further implications in very diverse fields, as genetics~\cite{A10} or finance~\cite{A10,T02}. 
 
In the original Parrondo's setup, the system under consideration had to show some degree of spatial inhomogeneity~\cite{HA99}, but further developments (partially) avoided this requirement by the inclusion of memory~\cite{PHA00,MB02,KJ03}, sometimes in a sophisticated (non-Markovian) way~\cite{CB02}. In our case, by contrast, we have reproduced the Parrondo's effect by means of a single, one-dimensional Markovian jump process, what represents up to our knowledge the first appearance of the paradox within the context of the CTRW.

The article is organized as follows: In Sec.~\ref{Sec_memoryless} we recall the traditional CTRW formalism based on the assumption of the independence between events. In Sec.~\ref{Sec_memory} we outline the fundamentals of the CTRWs with memory introduced in Ref.~\cite{MM07}, and devote a special attention to the case in which the correlation only affects the sign of consecutive jumps. Section \ref{Sec_alt} contains the main contribution of the paper, the proof that within the framework of the CTRWs with memory the mixture of two negatively biased processes may lead to a positively biased one. 
The paper ends with Sec.~\ref{Sec_Conc} where conclusions are drawn, and future perspectives are sketched.

\section{CTRWs without memory}
\label{Sec_memoryless}

Let us begin with a short review of the theory of CTRWs, for a more detailed explanation see e.g.~\cite{MMPW06}. The CTRW $X_a(t)$ is a stochastic process that, at random instants of time, $0=t_0\leq t_1\leq \cdots\leq t_{n-1}\leq t_n$, suffers random changes or {\it jumps\/} of magnitude $J_n$, 
\begin{equation*}
X_a(t)=\sum_{n=1}^{\infty}J_n \theta(t-t_n),
\end{equation*}
where  $\theta(u)=1$ for $u\geq 0$, and zero otherwise. 
In the simplest version of the formalism, the time intervals between consecutive changes, $\tau_n\equiv t_n-t_{n-1}$, and the random jumps are independent and identically distributed 
random variables, characterized by their corresponding probability density functions (PDFs) $\psi_a(\cdot)$ and $h_a(\cdot)$, 
\begin{eqnarray*}
\psi_a(\tau)d\tau &\equiv&
\PP\left\{\tau<\tau_n\leq \tau+d\tau\right\},\\
h_a(\x)d\x &\equiv&
\PP\left\{\x<J_n\leq \x+d\x\right\}.
\end{eqnarray*}
Henceforth we will use either $X_a(t)$ or the term ``process $A$'' whenever we want to refer to a CTRW that satisfies these requirements.

Let us introduce now the propagator $p_a(x,t)$, the transition probability of the process,
\begin{eqnarray*}
p_a(x,t)dx&\equiv&
\PP\left\{x<X_a(t+t_n)-X_a(t_n)\leq x+dx\right\}\\
&=&\PP\left\{x<X_a(t)\leq x+dx\right\},
\end{eqnarray*}
which follows a renewal equation~\cite{C65},
\begin{eqnarray}
p_a(x,t)=\delta(x)\int_{t}^{\infty} \psi_a(t') dt' \nonumber \\
+\int_{0}^{t}dt' \psi_a(t') \int_{-\infty}^{+\infty} h_a(\x) p_a(x-\x,t-t') d\x,
\label{renewal_pa}
\end{eqnarray}
thanks to the spatial and temporal invariance that the i.i.d. assumption brings to the problem |at least just after a jump.
It is well known that one can solve in a straight way Eq.~(\ref{renewal_pa}) for any choice of $\psi_a(\cdot)$ and $h_a(\cdot)$ in the Fourier-Laplace domain:
\begin{equation}
\hat{\tilde{p}}_a(\omega,s)=\frac{1-\hat{\psi}_a(s)}{s}\frac{1}{1-\hat{\psi}_a(s)\tilde{h}_a(\omega)},
\label{FL_pa}
\end{equation}
where here and hereafter the hat and/or the tilde over a function denotes its Laplace and/or Fourier transform with respect to its time and/or space variable, e.g.
\begin{equation*}
\hat{\psi}_a(s)\equiv\int_{0}^{\infty} \psi_a(t)e^{-s t} dt,
\end{equation*}
\begin{equation*}
\tilde{h}_a(\omega)\equiv\int_{-\infty}^{+\infty} h_a(\x)e^{i \omega \x} d\x,
\end{equation*}
and
\begin{equation*}
\hat{\tilde{p}}_a(\omega,s)\equiv\int_{0}^{\infty} dt e^{-s t}\int_{-\infty}^{+\infty} p_a(x,t)e^{i \omega x} dx.
\end{equation*}
Then, the computation of $p_a(x,t)$ reduces to the inversion of Eq.~(\ref{FL_pa}). To do this, however, the functional forms of $\psi_a(\cdot)$ and $h_a(\cdot)$ must be given. 

We will consider the issue of the waiting-time density in the first place: A CTRW cannot be a Markov process unless the time intervals between jumps are exponentially distributed. Such a lack of Markovianity would affect for instance the definition of the propagator and the scope of validity of Eq.~(\ref{renewal_pa}). The result is still tractable but, for a matter of simplicity in the modeling (and in the mixing) of the processes, we will simply consider that the number of changes (irrespective of the CTRW under consideration) is Poisson distributed, i.e., that the waiting-time PDF is always
\begin{equation*}
\psi(t)=\lambda e^{-\lambda t}.
\end{equation*}

The choice of $h_a(\cdot)$ is a far less delicate question. We could proceed without specifying the functional form of the jump-size distribution, but this would obscure most of the expressions in the forthcoming Sections, in particular the intermediate results. Thus, to prevent some key aspects from being buried by the mathematical terminology, we have decided to sacrifice a bit of generality for the sake of clarity, but without falling into an extreme simplicity: we have chosen that the jump sizes follow an asymmetric double exponential law,~\footnote{Due to their mathematical convenience, asymmetric double exponential jump distributions are commonly used in the modeling of financial processes with jumps, see e.g.~\cite{KW04,M08}.} 
\begin{equation}
h_a(\x)=q_0 \gamma_0 e^{-\gamma_0 \x} \theta(\x) +(1-q_0) \eta_0 e^{\eta_0 \x} \left[1- \theta(\x)\right],
\label{h_a}
\end{equation}
where the parameter $q_0\in[0,1]$ gives us the probability of having a non-negative jump, whereas $\gamma_0>0$ and $\eta_0>0$ are the inverses of the mean values of the absolute jump sizes in the upward and downward direction, respectively.

After all the above premises, Eq.~(\ref{FL_pa}) reads
\begin{equation*}
\hat{\tilde{p}}_a(\omega,s)=\frac{1}{s+\lambda\left[1-q_0\frac{\gamma_0}{\gamma_0-i\omega}-(1-q_0)\frac{\eta_0}{\eta_0+i\omega}\right]}.
\end{equation*}
The explicit knowledge of the previous characteristic function allows us to compute the mean value of the process, $\mu_a(t)$, in a simply way, since
\begin{equation*}
\hat{\mu}_a(s)=-i \left.\frac{\partial}{\partial \omega}\hat{\tilde{p}}_a(\omega,s)\right|_{\omega=0}=\left(\frac{q_0}{\gamma_0}-\frac{1-q_0}{\eta_0}\right)\frac{\lambda}{s^2},
\end{equation*}
and finally, after the Laplace inversion
\begin{equation}
\mu_a(t)= \left(\frac{q_0}{\gamma_0}-\frac{1-q_0}{\eta_0}\right)\lambda t=\mu_0 \lambda t,
\label{mu_a}
\end{equation}
where we have defined
\begin{equation}
\mu_0\equiv \frac{q_0}{\gamma_0}-\frac{1-q_0}{\eta_0},
\label{mu_0}
\end{equation}
which is just the mean value of the jump sizes,
\begin{equation}
\mu_0= \int_{-\infty}^{+\infty} \x h_a(\x) d\x.
\end{equation}
In fact, one can easily prove that this result is valid for any choice of $h_a(\cdot)$, provided that the mean value $\mu_0$ does exist.

Returning to the case we are analyzing in detail, Eq.~(\ref{mu_0}) shows that one will obtain an unbiased process ($\mu_0=0$) whenever one imposes on the positive parameters $q_0$, $\gamma_0$, and $\eta_0$ the constraint
\begin{equation}
q_0=\frac{\gamma_0}{\gamma_0+\eta_0}, 
\label{unbiased_a}
\end{equation}
which is always feasible.

\section{CTRWs with memory}
\label{Sec_memory}

We will consider now a second process, the process $B$, 
\begin{equation*}
X_b(t)=\sum_{n=1}^{\infty}\mathcal{J}_n \theta(t-t_n),
\end{equation*}
a CTRW that belongs to the class of processes introduced in~\cite{MM07}. The main idea behind the work developed in this reference is the following: one can introduce memory effects into the framework of the CTRW without paying a too high price by demanding to the jump sizes $\mathcal{J}_n$ that satisfy the Markov property,
\begin{eqnarray*}
& &\PP\left\{\left.\x<\mathcal{J}_n\leq \x+d\x\right|\mathcal{J}_{n-1}=\y,\dots,\mathcal{J}_{0}=\z\right\}\\
&=&\PP\left\{\left.\x<\mathcal{J}_n\leq \x+d\x\right|\mathcal{J}_{n-1}=\y\right\}\equiv h_b(\x|\y) d\x.
\end{eqnarray*}
In such a case the process itself is not Markovian, but even then one can derive renewal equations for the propagator, conditioned to the last-known jump value,
\begin{eqnarray*}
p_b(x,t|\y)dx\equiv
\PP\left\{\left.x<X_b(t)\leq x+dx\right|\mathcal{J}_0=\y\right\}.
\end{eqnarray*}

Perhaps the simplest approach that one can adopt to introduce this kind of short-ranged memory is through persistence~\cite{CB02,VP07,HV10}, i.e. that the dependence of the process on the previous history is restricted to the last-jump sign:
\begin{equation}
h_b(\x|\y)=h_1(\x)\theta(\y)+h_2(\x)\left[1- \theta(\y)\right],
\label{h_b}
\end{equation}
a model that, in spite of its simplicity, is of applied interest. For instance, there are evidences that suggest that a model of this kind is precise enough to describe with good accuracy the behavior of highly traded equities~\cite{MPMLMM05}.

For the same reasons given in the previous Section,
we will assume here that the conditional distribution of the jumps are two asymmetric double exponential functions:
\begin{eqnarray*}
h_{1}(\x)&=&q_{1} \gamma_{1} e^{-\gamma_{1} \x} \theta(\x) +(1-q_{1}) \eta_{1} e^{\eta_{1} \x} \left[1- \theta(\x)\right],\\
h_{2}(\x)&=&q_{2} \gamma_{2} e^{-\gamma_{2} \x} \theta(\x) +(1-q_{2}) \eta_{2} e^{\eta_{2} \x} \left[1- \theta(\x)\right],
\end{eqnarray*}
with $q_{1,2}\in [0,1]$, and $\gamma_1$, $\eta_1$, $\gamma_2$, and $\eta_2$ positive.

In concordance with~(\ref{h_b}), the conditional propagator $p_b(x,t|\y)$ can be reduced to two different functions depending on whether the previous change of the process had a positive sign, $p_b(x,t|+)$, or a negative sign, $p_b(x,t|-)$.~\footnote{We can include the zero-amplitude jump within the positive case, or just ignore it since it is of null measure.} As we have stated above, we can write down renewal equations for these magnitudes: specifically a set of two coupled Volterra integral equations of the second kind,
\begin{eqnarray}
&&p_b(x,t|+)=\delta(x)e^{-\lambda t} +\int_{0}^{t}dt' \lambda e^{-\lambda t'}\nonumber \\&\times&\left\{q_1 \int_{0}^{+\infty} \gamma_1 e^{-\gamma_1 \x} p_b(x-\x,t-t'|+) d\x\right.\nonumber \\ &+&\left.(1-q_1)\int_{-\infty}^{0} \eta_1 e^{\eta_1 \x} p_b(x-\x,t-t'|-) d\x\right\},
\label{renewal_pbp}
\end{eqnarray}
and
\begin{eqnarray}
&&p_b(x,t|-)=\delta(x)e^{-\lambda t}+ \int_{0}^{t}dt' \lambda e^{-\lambda t'}\nonumber \\&\times&\left\{q_2 \int_{0}^{+\infty} \gamma_2 e^{-\gamma_2 \x} p_b(x-\x,t-t'|+) d\x\right.\nonumber \\ &+&\left.(1-q_2)\int_{-\infty}^{0} \eta_2 e^{\eta_2 \x} p_b(x-\x,t-t'|-) d\x\right\}.
\label{renewal_pbm}
\end{eqnarray}
We may explain how Eqs.~(\ref{renewal_pbp}) and~(\ref{renewal_pbm}) were derived by analyzing their common three-piece structure.  The first term in both expressions takes into account the possibility that the process remains unchanged throughout the time interval $t$. In the second and third terms at least an event occurred at time $t'$, $0\leq t' \leq t$: the difference among these two contributions comes from the fact that, in the second term the jump was upward, $\x>0$, and the process renews from that point, $p_b(x-\x,t-t'|+)$, whereas in the third term the jump was downward, $\x<0$, and therefore the subsequent propagator is $p_b(x-\x,t-t'|-)$. Note that Eq.~(\ref{renewal_pbp}) will differ from Eq.~(\ref{renewal_pbm}) as long as $h_1(\cdot)$ is not coincident with $h_2(\cdot)$.

To solve the posed problem we will resort back to the Fourier-Laplace transform. The two integral equations~(\ref{renewal_pbp}) and~(\ref{renewal_pbm}) turn into a set of two algebraic equations when moved into the Fourier-Laplace domain,
\begin{eqnarray*}
\hat{\tilde{p}}_b(\omega,s|+)&=&\frac{1}{\lambda+s}\nonumber \\&+&\frac{\lambda}{\lambda+s}\left\{q_1\frac{\gamma_1}{\gamma_1-i\omega}\hat{\tilde{p}}_b(\omega,s|+)\right.\nonumber \\ &+&\left.(1-q_1)\frac{\eta_1}{\eta_1+i\omega} \hat{\tilde{p}}_b(\omega,s|-) \right\},
\end{eqnarray*}
and
\begin{eqnarray*}
\hat{\tilde{p}}_b(\omega,s|-)&=&\frac{1}{\lambda+s}\nonumber \\&+&\frac{\lambda}{\lambda+s}\left\{q_2\frac{\gamma_2}{\gamma_2-i\omega}\hat{\tilde{p}}_b(\omega,s|+)\right.\nonumber \\ &+&\left.(1-q_2)\frac{\eta_2}{\eta_2+i\omega} \hat{\tilde{p}}_b(\omega,s|-) \right\},
\end{eqnarray*}
whose solution reads
\begin{eqnarray}
\hat{\tilde{p}}_b(\omega,s|+)&=&
\frac{s+\lambda \left[1+\frac{(1-q_1)\eta_1}{\eta_1+i\omega}-\frac{(1-q_2)\eta_2}{\eta_2+i\omega}\right]}{\hat{\tilde{\Delta}}_b(\omega,s)},
\label{FL_pbp}
\\
\hat{\tilde{p}}_b(\omega,s|-)&=&
\frac{s+\lambda \left[1-\frac{q_1\gamma_1}{\gamma_1-i\omega}+\frac{q_2\gamma_2}{\gamma_2-i\omega}\right]}{\hat{\tilde{\Delta}}_b(\omega,s)},
\label{FL_pbm}
\end{eqnarray}
with
\begin{eqnarray}
\hat{\tilde{\Delta}}_b(\omega,s)&=&
\left(s+\lambda -\frac{\lambda q_1\gamma_1}{\gamma_1-i\omega}\right)\left(s+\lambda-\frac{\lambda(1-q_2)\eta_2}{\eta_2+i\omega}\right)
\nonumber\\
&-&
\lambda^2(1-q_1)q_2 \frac{\gamma_2}{\gamma_2-i\omega}\frac{\eta_1}{\eta_1+i\omega}
.
\label{FL_Deltab}
\end{eqnarray}
We can now compute the {\it unconditional\/} transition PDF, 
\begin{eqnarray}
\hat{\tilde{p}}_b(\omega,s)=\beta\hat{\tilde{p}}_b(\omega,s|+)+(1-\beta)\hat{\tilde{p}}_b(\omega,s|-),
\label{FL_pb}
\end{eqnarray}
where $\beta$ is the likelihood that a given jump takes the positive sign, which follows from the total probability theorem:
\begin{equation}
\beta=\beta q_1 +(1-\beta) q_2 \Rightarrow \beta=\frac{q_2}{1-q_1+q_2}.
\label{beta}
\end{equation}
The unconditional mean value of the process can be obtained from the differentiation of (\ref{FL_pb}) with respect to $\omega$, for $\omega=0$, and reads in the Laplace domain:
\begin{eqnarray}
\hat{\mu}_b(s)&=&\frac{\lambda}{s^2}\left[\beta\left(\frac{q_1}{\gamma_1}-\frac{1-q_1}{\eta_1}\right)+(1-\beta)\left(\frac{q_2}{\gamma_2}-\frac{1-q_2}{\eta_2}\right)\right]\nonumber \\&=& \frac{\lambda}{s^2}\left[\beta \mu_1+(1-\beta)\mu_2\right],
\label{L_mub}
\end{eqnarray}
where we have denoted by $\mu_1$ and $\mu_2$ the first moments of the PDFs $h_1(\cdot)$ and $h_2(\cdot)$ respectively,
\begin{eqnarray*}
\mu_1&\equiv&\frac{q_1}{\gamma_1}-\frac{1-q_1}{\eta_1},\\
\mu_2&\equiv&\frac{q_2}{\gamma_2}-\frac{1-q_2}{\eta_2}.
\end{eqnarray*}
The inverse Laplace transform of~(\ref{L_mub}) is straightforward and yields 
\begin{eqnarray}
\mu_b(t)
&=&\left[\beta \mu_1+(1-\beta)\mu_2\right]\lambda t,
\label{mu_b}
\end{eqnarray}
and therefore process $B$ will become unbiased whenever
\begin{equation}
\beta \mu_1+(1-\beta)\mu_2=0.
\label{unbiased_b}
\end{equation}
Like in~(\ref{unbiased_a}), condition~(\ref{unbiased_b}) can be met by selecting $q_1$ and/or $q_2$ in a proper way. For instance
\begin{equation*}
q_2=\frac{1}{1+\eta_2\left(\frac{1}{\gamma_2}-\frac{1}{\eta_1}+\frac{1}{\gamma_1}\frac{q_1}{1-q_1}\right)}
\end{equation*}
will remove the bias from process $B$ for any choice of $q_1$, $\gamma_1$, $\eta_1$, $\gamma_2$ and $\eta_2$, provided that $\gamma_2\leq\eta_1$. If $\gamma_2>\eta_1$, one must request $q_1$ to satisfy the supplementary constraint:
\begin{equation*}
q_1>\frac{\frac{1}{\eta_1}-\frac{1}{\gamma_2}}{\frac{1}{\eta_1}-\frac{1}{\gamma_2}+\frac{1}{\gamma_1}}.
\end{equation*}
 
Let us conclude the analysis of process $B$ by noting that the result in~(\ref{mu_b}) is more general than what it could be presumed. The expression is still valid if one considers any alternative for the PDFs $h_1(\cdot)$ and $h_2(\cdot)$ in~(\ref{h_b}), with the sole condition that their mean values $\mu_1$ and $\mu_2$ are bounded. One must be aware, however, that expression~(\ref{beta}) for $\beta$ 
should be recomputed.

\section{Alternation of processes}
\label{Sec_alt}

In this Section we will analyze the outcome of the combination of the two previous processes, $A$ and $B$, what will lead to the new process $AB$, a process that also belongs to the class of CTRWs with memory. To do this, we will assume that the mixing procedure is random: we will have a probability $r$ that the next process increment follows Eq.~(\ref{h_a}) and $1-r$ that the change is driven by Eq.~(\ref{h_b}).~\footnote{When similar problems have been analyzed in the context of game theory, by ``process $AB$'' one may refer to the {\it deterministic} alternation of the two games. This is not the case here.} The renewal equations for the conditional propagators have a structure that is very similar to that in Eqs.~(\ref{renewal_pbp}) and (\ref{renewal_pbm}), but where $h_b(\x|\y)$ has been replaced by
\begin{equation}
h(\x|\y)=r h_a(\x)+(1-r)h_b(\x|\y),
\label{h_ab}
\end{equation}
namely
\begin{eqnarray}
p(x,t|+)&=&\delta(x)e^{-\lambda t} +\int_{0}^{t}dt' \lambda e^{-\lambda t'}
\nonumber \\&\times&
\left\{\int_{0}^{+\infty}d\x K_{++}(\x) p(x-\x,t-t'|+)  \right.
\nonumber \\
&+&\left.\int_{-\infty}^{0}d\x K_{-+}(\x) p(x-\x,t-t'|-)\right\},
\nonumber 
\end{eqnarray}
\begin{eqnarray}
p(x,t|-)&=&\delta(x)e^{-\lambda t} +\int_{0}^{t}dt' \lambda e^{-\lambda t'}
\nonumber \\ &\times&
\left\{\int_{0}^{+\infty}d\x K_{+-}(\x) p(x-\x,t-t'|+)  \right.
\nonumber \\
&+&\left.\int_{-\infty}^{0}d\x K_{--}(\x) p(x-\x,t-t'|-)\right\},
\nonumber
\end{eqnarray}
with
\begin{eqnarray*}
K_{++}(\x)&=&r q_0  \gamma_0 e^{-\gamma_0 \x}+ (1-r) q_1  \gamma_1 e^{-\gamma_1 \x},\\
K_{-+}(\x)&=&r (1-q_0)  \eta_0 e^{\eta_0 \x}+ (1-r)(1- q_1)  \eta_1 e^{\eta_1 \x},\\
K_{+-}(\x)&=&r q_0  \gamma_0 e^{-\gamma_0 \x}+ (1-r) q_2  \gamma_2 e^{-\gamma_2 \x},\\
K_{--}(\x)&=&r (1-q_0)  \eta_0 e^{\eta_0 \x}+ (1-r)(1- q_2)  \eta_2 e^{\eta_2 \x}.
\end{eqnarray*}

The solution of the new posed problem in the Fourier-Laplace domain also mimics Eqs.~(\ref{FL_pbp})|(\ref{FL_Deltab}),
\begin{eqnarray*}
\hat{\tilde{p}}(\omega,s|+)&=&
\frac{s+\lambda \left\{1+(1-r)\left[\frac{(1-q_1)\eta_1}{\eta_1+i\omega}-\frac{(1-q_2)\eta_2}{\eta_2+i\omega}\right]\right\}}{\hat{\tilde{\Delta}}(\omega,s)},
\\
\hat{\tilde{p}}(\omega,s|-)&=&
\frac{s+\lambda \left\{1-(1-r)\left[\frac{q_1\gamma_1}{\gamma_1-i\omega}-\frac{q_2\gamma_2}{\gamma_2-i\omega}\right]\right\}}{\hat{\tilde{\Delta}}(\omega,s)},
\end{eqnarray*}
and
\begin{eqnarray*}
\hat{\tilde{\Delta}}(\omega,s)&=&%
\lambda^2\left[\frac{s}{\lambda}+1 -\frac{ r q_0\gamma_0}{\gamma_0-i\omega}-\frac{ (1-r) q_1\gamma_1}{\gamma_1-i\omega}\right]\nonumber \\
&\times&\left[\frac{s}{\lambda}+1-\frac{ r(1-q_0)\eta_0}{\eta_0+i\omega}-\frac{(1-r)(1-q_2)\eta_2}{\eta_2+i\omega}\right]\nonumber\\
&-&\lambda^2\left[ \frac{r q_0\gamma_0}{\gamma_0-i\omega}+\frac{(1-r) q_2\gamma_2}{\gamma_2-i\omega}\right]\nonumber \\
&\times&\left[\frac{r(1-q_0)\eta_0}{\eta_0+i\omega}+\frac{(1-r)(1-q_1)\eta_1}{\eta_1+i\omega}\right].
\end{eqnarray*}

\begin{figure}[hbtp] 
\hfil \includegraphics[width=0.90\columnwidth,keepaspectratio=true]{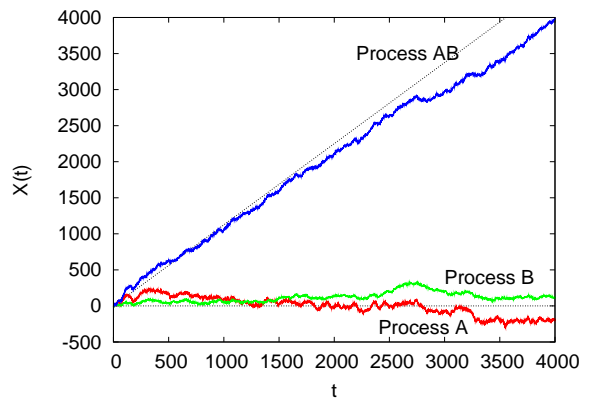}
\caption{(Color online) Sample paths of the processes analyzed in the text. The parameter values are $\lambda=20$, $q_0=1/2$, $q_1=q_2=4/5$, $\gamma_0=\eta_0=\eta_1=\gamma_2=\eta_2=1$, $\gamma_1=16$, and $r=1/2$, what renders $\mu_a(t)=\mu_b(t)=0$, and $\mu(t)=9 t/8$, represented in the plot by (black) dotted lines.}
\label{Fig1}
\end{figure}

As in the previous Section, we can recover the unconditional propagator $\hat{\tilde{p}}(\omega,s)$ by means of $\hat{\tilde{p}}(\omega,s|\pm)$,  
\begin{eqnarray}
\hat{\tilde{p}}(\omega,s)=\alpha\hat{\tilde{p}}(\omega,s|+)+(1-\alpha)\hat{\tilde{p}}(\omega,s|-),
\label{FL_p}
\end{eqnarray}
and  the {\it new\/} stationary probability of having a positive change,  $\alpha$. This probability is now the result of the combined effect of the two individual processes:
\begin{eqnarray}
\alpha&=&r q_0 +(1-r)\left[\alpha q_1 +(1-\alpha) q_2\right] \Rightarrow \nonumber \\
\alpha&=&\frac{r q_0 + (1-r) q_2}{1-(1-r)(q_1-q_2)}.
\label{alpha}
\end{eqnarray}
One more time we will differentiate Eq.~(\ref{FL_p}) and obtain eventually the unconditional mean value of the process $AB$,
\begin{eqnarray}
\mu(t)=r \mu_a(t)+(1-r)\left[\alpha\mu_1+(1-\alpha)\mu_2\right]\lambda t,
\label{mu}
\end{eqnarray}
which is {\it not\/} the linear superposition of the individual mean values of processes $A$ and $B$, 
\begin{equation}
\mu(t)\neq r \mu_a(t)+(1-r)\mu_b(t),
\end{equation}
unless $\alpha=\beta$. The ultimate origin of this non-linearity lies in the fact that, as Eq.~(\ref{h_ab}) shows, the correlation of process $B$ is affected by the inclusion of process $A$. Therefore, even in the case in which $\mu_a(t)=0$, $\mu_b(t)=0$, the composite process will exhibit a clear drift if $\alpha\neq\beta$, as it can be seen in Fig.~\ref{Fig1}, where we plot a possible realization of processes $A$, $B$, and $AB$.   

Since $\alpha$ is a function of $r$, we can tune this parameter to amplify the paradoxical effect. Let us search the value of $r$ for which the drift is maximum, by direct differentiation of Eq.~(\ref{mu}), under the assumptions (\ref{unbiased_a}) and (\ref{unbiased_b}):
\begin{eqnarray}
\frac{\partial\mu(t)}{\partial r}=-\frac{\left[q_0 \mu_1 +(1-q_0)\mu_2 \right](r-r_+)(r-r_-)}{\left[1-(1-r)(q_1-q_2)\right]^2}\lambda t,
\label{partial_u}
\end{eqnarray}
with
\begin{equation*}
r_{\pm}=\frac{\sqrt{1-q_1+q_2}\pm\left(1-q_1+q_2\right)}{q_1-q_2}.
\end{equation*}
Equation~(\ref{partial_u}) has three possible zeros. The first one corresponds to 
\begin{equation*}
q_0 \mu_1 +(1-q_0)\mu_2 =0\Rightarrow q_0=\beta,
\end{equation*}
but in this case $\alpha=\beta$ as well, and $\mu(t)=0$, irrespective of $r$. The second one, $r=r_+$,
is not valid because, as it can be shown, it is either smaller than zero or greater than 1. The last one will provide us with the optimal drift enhancement, $r=r_-$. The value of $r_-$ can always be interpreted as a mixing probability, as it fulfills $r_-\in[0,1]$ for any given choice of $q_1$ and $q_2$, and gives $r_-=1/2$ for $q_1=q_2$. This is just the case considered in 
Fig.~\ref{Fig1}. 

\begin{figure}[hbtp] 
\hfil \includegraphics[width=0.90\columnwidth,keepaspectratio=true]{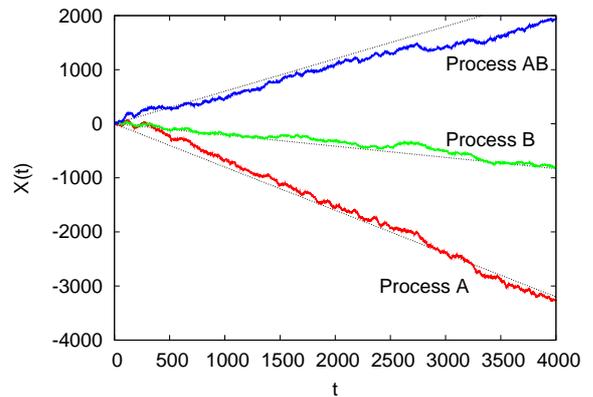}
\caption{(Color online) Sample paths of the biased processes. The parameters coincide with those in Fig.~\ref{Fig1}, except that $q_0$, $q_1$, and $q_2$ were diminished by the same quantity $\epsilon=0.02$, i.e. $q_0=12/25$, $q_1=q_2=39/50$. The (black) dotted lines show the mean value of each process.}
\label{Fig2}
\end{figure}

Now, once we have shown that the interaction of two unbiased processes can bring a new process with positive drift, it is not difficult to reduce the probabilities $q_0$, $q_1$ and $q_2$ by a small quantity $\epsilon$, in such a way the mean value of anyone of the two individual processes is negative, but the combined process presents a positive growth. If we use as a starting point the values reported in the caption of Fig.~\ref{Fig1} above, that is $q_0=1/2-\epsilon$, $q_1=q_2=4/5-\epsilon$, $\gamma_0=\eta_0=\eta_1=\gamma_2=\eta_2=1$, $\gamma_1=16$, and $r=1/2$, we will have that
\begin{eqnarray*}
\mu_a(t)&=&-2\epsilon \lambda t<0, \\
\mu_b(t)&=&-\frac{8\epsilon +15 \epsilon^2}{16}\lambda t<0,\\
\mu(t)&=&\frac{36-833\epsilon-340\epsilon^2}{640} \lambda t.
\end{eqnarray*}
The last expression is positive for $\epsilon\lesssim 0.0425$. Therefore, if we set $\epsilon=0.02$, as in Fig.~\ref{Fig2}, we will achieve the desired behavior, that the bias of process $AB$ is in the opposite direction of those of process $A$ and process $B$.

Note however that we can examine the physical meaning of process $AB$ from another interesting perspective: we may understand that process $A$ is a noise source affecting process $B$. This identification is even more natural when $q_0=1/2$ and $\gamma_0=\eta_0$, as in Fig.~\ref{Fig1} above, because then the increments of process $A$ constitute a zero-mean, symmetric white noise. In that case, one can see how the intensification of the noise steadily increases the mean output of the process, until it reaches a maximum, like in the case of the stochastic resonance. 

\begin{figure}[hbtp] 
\hfil \includegraphics[width=0.90\columnwidth,keepaspectratio=true]{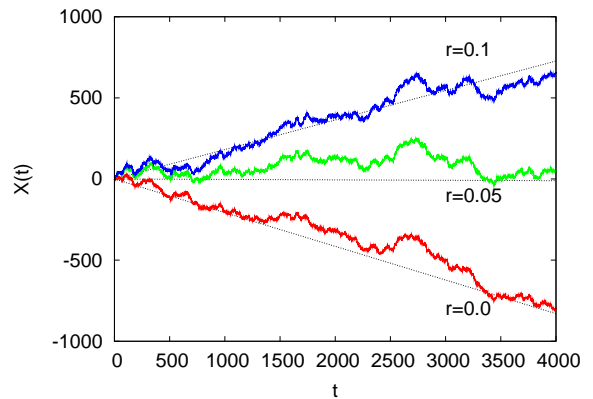}
\caption{(Color online) Sample paths of the process $AB$ with variable bias. The parameters are $\lambda=20$, $q_0=1/2$, $q_1=q_2=39/50$, $\gamma_0=\eta_0=\eta_1=\gamma_2=\eta_2=1$, and $\gamma_1=16$. With this setup $\mu_a(t)=0$, $\mu_b(t)<0$ and the sign of $\mu(t)$ depends on the value of the noise level $r$, as it is shown by the superimposed (black) dotted lines.}
\label{Fig3}
\end{figure}

But we can go further in the exploration of the possible implications of the paradoxical behavior. Let us consider the following values for the parameters of our example: $q_0=1/2$, $q_1=q_2=39/50$, $\gamma_0=\eta_0=\eta_1=\gamma_2=\eta_2=1$, and $\gamma_1=16$. The corresponding mean values of the three processes, now as a function of $r$, are
\begin{eqnarray*}
\mu_a(t)&=&0, \\
\mu_b(t)&=&-\frac{83}{8000}\lambda t<0,\\
\mu(t)&=&(1-r)\frac{1638 r-83}{8000} \lambda t,
\end{eqnarray*}
and therefore $\mu(t)>0$ for $r>83/1638\approx 0.05$. This represents that we can modify the sign of the output of the system $AB$ by changing the level of the noise, as it is shown in Fig.~\ref{Fig3}. This broadens the fields for which one may find practical applications of this formalism: from game theory to stochastic control.

\section{Conclusions}
\label{Sec_Conc}

We have shown with a 
particular but illuminating example how we can obtain a growing stochastic process by alternating two unbiased CTRWs, one of them with memory. The clue to the understanding of this effect is in the fact that the mixing of the two processes distorts the inner correlation of the CTRW with memory.
 
The phenomenon is related to the Parrondo's paradox in game theory where the alternative play of two losing games may give winnings to the player. In our case, we can modify the parameters controlling the two CTRWs in such a way that each separate process acquires a negative drift but that their interplay still produces a positive bias.

The peculiarities of the analyzed process and its noticeable connections with the outcome of a gambling game do not limit the scope of our results, however. 
In the first place, the assumption that the only relevant past information is confined into the jump signs does not seem to be essential for obtaining the Parrondo's effect: we have selected it because it is convenient from a mathematical point of view, given its simplicity, and proves to be a plausible mechanism for modeling actual systems, at least in the realm of finance. And in the second place, as in the case of the original analysis by Parrondo, the paradox has consequences on problems far away from game theory: in our case we have shown how to control the sign of the (average value of the) output of a system by increasing or decreasing the intensity of a noise source. 

Therefore, we think that the search for the emergence of this Parrondo-like behavior in CTRWs with more sophisticated memory functions, and the suggestion of alternative interpretations of the paradox in these new contexts, can be very fruitful from both the theoretical and the applied point of view. In any case, this will be the matter of a future work.

\acknowledgments

The author wishes to thank the anonymous referees for their comments, and acknowledges support from the Spanish {\it Ministerio de Ciencia e Innovaci\'on\/} under contract No. FIS2009-09689, and from {\it Generalitat de Catalunya\/}, contract No. 2009SGR417.


\begin{thebibliography}{00}
\bibitem{MW65} \Journal{E. W. Montroll and G. H. Weiss}{1965}{Random Walks on Lattices, II}{J. Math. Phys.}{6}{167--181}

\bibitem{W94} \Book{G. H. Weiss}{1994}{Aspects and Applications of the Random Walk}{North-Holland}{Amsterdam}

\bibitem{S74} \Journal{M. F. Shlesinger}{1974}{} {J. Stat. Phys.}{10}{421-434}%

\bibitem{MS84} \Book{E. W. Montroll and M. F. Shlesinger}{1984}{
{\rm in} Nonequilibrium Phenomena II: From Stochastics to Hydrodynamics, {\rm edited by J. L. Lebowitz and E. W. Montroll, pp. 1--121}}{North-Holland}{Amsterdam}
  
\bibitem{WPM98} \Journal{G. H. Weiss, J. M. Porr\`a, and J. Masoliver}{1998}{ }{Phys. Rev. E}{58}{6431--6439}

\bibitem{MaBe02} \Journal{G. Margolin and B. Berkowitz}{2002}{}{Phys. Rev. E}{65}{031101}


\bibitem{HWS82} \Journal{B. D. Hughes,  E. W.  Montroll, and  M. F. Shlesinger}{1982}{Fractal random walks} {J. Stat. Phys.}{28}{111-126}%

\bibitem{GW02} \Journal{E. Gudowska-Nowak and K. Weron}{2002}{}{Phys. Rev. E}{65}{011103}

\bibitem{BC97} \Journal{M. Bogu\~n\'a and \'A. Corral}{1997}{}{Phys. Rev. Lett.}{78}{4950--4953}

\bibitem{HS02} \Journal{A. Helmstetter and D. Sornette}{2002}{}{Phys. Rev. E}{66}{061104}

\bibitem{MAGLP03} \Journal{M. S. Mega, P. Allegrini, P. Grigolini, V. Latora, and  L. Palatella}{2003}{}{Phys. Rev. Lett.}{90}{188501}

\bibitem{C06} \Journal{\'A. Corral}{2006}{}{Phys. Rev. Lett.}{97}{178501}


\bibitem{BS97} \Journal{B. Berkowitz and H. Scher}{1997}{}{Phys. Rev. Lett.}{79}{4038--4041}

\bibitem{DWG03} \Journal{L. Dagdug, G. H.  Weiss, and A. H. Gandjbakhche}{2003}{}{Phys. Med. Biol.}{48}{1361--1370}

\bibitem{DW05} \Journal{O. K.  Dudko and  G. H.  Weiss}{2005}{ }{Diff. Fund.}{2}{1--21}

\bibitem{BKMS01} \Journal{B. Berkowitz, G. Kosakowski, G. Margolin, and H. Sher}{2001}{}{Ground Water}{39}{593--604}

\bibitem{DB03} \Journal{M. Dentz and B. Berkowitz}{2003}{}{Water Resources Research}{39}{1111}

\bibitem{SGM00} \Journal{E. Scalas, R. Gorenflo, and F. Mainardi}{2000}{}{Physica A}{284}{376--384}

\bibitem{MMW03} \Journal{J. Masoliver, M. Montero, and G. H. Weiss}{2003}{}{Phys. Rev. E}{67}{021112}

\bibitem{KS03} \Journal{R. Kutner and F. Swita{\l}a}{2003}{}{Quantitative Finance}{3}{201--211}

\bibitem{RR04} \Journal{P. Repetowicz and P. Richmond}{2004}{}{Physica A}{344}{108--111}

\bibitem{MMP05} \Journal{J. Masoliver, M. Montero, and J. Perell\'o}{2005}{}{Phys. Rev. E}{71}{056130}

\bibitem{MPMLMM05} \Journal{M. Montero, J.  Perell\'o, J. Masoliver,  F. Lillo, S. Miccich\'{e}, and R. N. Mantegna}{2005}{}{Phys. Rev. E}{72}{056101}

\bibitem{S06} \Journal{E. Scalas}{2006}{}{Physica A}{362}{225--239}

\bibitem{MMPW06} \Journal{J. Masoliver, M. Montero, J.  Perell\'o, and G. H. Weiss}{2006}{}{J. Econ. Behav. Organ.}{61}{577--598}

\bibitem{GPR01} \Journal{P. Grigolini, L. Palatella, and G. Raffaelli}{2001}{}{Fractals}{9}{439--449}

\bibitem{K02} \Journal{R. Kutner}{2002}{}{Chem. Phys.}{284}{481--505}

\bibitem{SGM04} \Journal{E. Scalas, R. Gorenflo, and F. Mainardi}{2004}{}{Phys. Rev. E}{69}{011107}

\bibitem{GPSL09} \Journal{G. Germano, M. Politi, E. Scalas, and R. L. Schilling}{2009}{}{Phys. Rev. E}{79}{066102}

\bibitem{C01} \Journal{R. Cont}{2001}{Empirical properties of asset returns: stylized facts and statistical issues}{Quantitative Finance}{1}{223--236}

\bibitem{GK10} \Journal{T. Gubiec and R. Kutner}{2010}{Backward jump continuous-time random walk: An application to market trading}{Phys. Rev. E}{82}{046119}

\bibitem{MM07} \Journal{M. Montero and J. Masoliver}{2007}{Nonindependent continuous-time random walks}{Phys. Rev. E}{76}{061115}

\bibitem{CM65} \Book{D. R. Cox and H. D. Miller}{1965}{The Theory of Stochastic Processes}{Wiley}{New York}{1965}



\bibitem{AAP05} \Book{A. Allison, D. Abbott, and C. Pearce}{2005}{
{\rm in} Advances in Dynamic Games: Applications to Economics, Finance, Optimization, and Stochastic Control, {\rm edited by A. S. Nowak and
K. Szajowski, pp. 613--633}}{Birkh\"auser}{Boston} 

\bibitem{HA99} \Journal{G. P. Harmer and D. Abbott}{1999}{Losing strategies can win by Parrondo's paradox}{Nature}{402}{864}

\bibitem{HAT00} \Journal{G. P. Harmer, D. Abbott, and P. G. Taylor}{2000}{The paradox of Parrondo's games}{Proc. R. Soc. A}{456}{247--259} 

\bibitem{AATA04} \Journal{P. Amengual, A. Allison, R. Toral, and D. Abbott}{2004}{Discrete-time ratchets, the Fokker--Planck equation and Parrondo's paradox}{Proc. R. Soc. A}{460}{2269--2284} 

\bibitem{A10} \Journal{D. Abbott}{2010}{Asymmetry and Disorder: A Decade of Parrondo's Paradox}{Fluct. Noise Lett.}{9}{129--156} 

\bibitem{T02} \Journal{R. Toral}{2002}{Capital Redistribution Brings Wealth By Parrondo'S Paradox}{Fluct. Noise Lett.}{2}{L305--L311}

\bibitem{PHA00} \Journal{J. M. R. Parrondo, G. P. Harmer, and D. Abbott}{2000}{New Paradoxical Games Based on Brownian Ratchets}{Phys. Rev. Lett.}{85}{5226--5229}

\bibitem{MB02} \Journal{D. A. Meyer and H. Blumer}{2002}{Parrondo Games as Lattice Gas Automata}{J. Stat. Phys.}{107}{225--239} 

\bibitem{KJ03} \Journal{R. J. Kay and N. F. Johnson}{2003}{Winning combinations of history-dependent games}{Phys. Rev. E}{67}{056128} 

\bibitem{CB02} \Journal{B. Cleuren and C. Van den Broeck}{2002}{Random walks with absolute negative mobility}{Phys. Rev. E}{65}{030101(R)} 

\bibitem{C65} \Book{D. R. Cox}{1965}{Renewal Theory}{John Wiley and Sons}{New York}

\bibitem{KW04} \Journal{S. G. Kou and H. Wang}{2004}{Option Pricing Under a Double Exponential Jump Diffusion Model}{Manage. Sci.}{50}{1178--1192}

\bibitem{M08} \Journal{M. Montero}{2008}{Renewal equations for option pricing}{Eur. Phys. J. B}{65}{295--306}

\bibitem{VP07} \Journal{P. Vallois and C. S. Tapiero}{2007}{Memory-based persistence in a counting random walk process}{Physica A}{386}{303--317} 

\bibitem{HV10} \Journal{S. Herrmann and P. Vallois}{2010}{From Persistent Random Walk To The Telegraph Noise}{Stoch. Dyn.}{10}{161--196} 
\end{thebibliography}
\end{document}